\documentstyle[preprint,pre,aps]{revtex}  
\tighten

\begin{document}
\draft

\title{Correlation functions in the two-dimensional\\  random-bond Ising model}
\author{
S. L. A. de Queiroz$^a$\footnote{Electronic address:
sldq@if.uff.br} and
 R. B. Stinchcombe$^b$\footnote{Electronic address: stinch@thphys.ox.ac.uk}
 }
\address{
$^a$ Instituto de F\'\i sica, Universidade Federal Fluminense,\\ Avenida
Litor\^anea s/n, Campus da Praia Vermelha, 24210-340 Niter\'oi RJ, Brazil \\
$^b$ Department of Physics, Theoretical Physics, University of Oxford,\\
 1 Keble Road, Oxford OX1 3NP, United Kingdom
}
\date{\today}
\maketitle
\begin{abstract}
We consider long strips of finite width $L \leq 13$ sites 
of ferromagnetic Ising spins
 with random couplings distributed  according to the binary distribution:
 $P(J_{ij})= {1 \over 2} ( \delta (J_{ij} -J_0) +  \delta (J_{ij} -rJ_0) ) 
,\ 0 < r < 1 $. Spin-spin
correlation functions 
$ <\sigma_{0} \sigma_{R}>$ along the ``infinite'' direction are computed by transfer-matrix methods, at the critical temperature of the corresponding
two-dimensional system, and their probability distribution is investigated.
We show that, although in-sample fluctuations do not die out as strip length is
increased, averaged values  converge satisfactorily. These latter are very close
to the critical correlation functions of the pure Ising model, in agreement with
recent Monte-Carlo simulations. A scaling approach is
formulated, which provides the essential aspects of the $R$-- and $L$--
dependence of the probability distribution of 
$\ln <\sigma_{0} \sigma_{R}>$, including the result that the
appropriate scaling variable is $R/L$. Predictions from scaling theory
are borne out by  numerical data, which show the
probability distribution of 
$\ln <\sigma_{0} \sigma_{R}>$ to be remarkably skewed at
short distances, approaching a Gaussian only as $R/L \gg 1$ .    
\end{abstract}

\narrowtext

\section{Introduction}

Most studies of random magnetic systems focus on whether or not quenched
disorder destroys a sharp phase transition and, in the latter case, whether critical exponents are  the same
as for the corresponding pure magnets~\cite{st83,sh94,sst}.
Less attention has been paid to the underlying 
probability distribution functions
which govern the behaviour of
sample-averaged thermodinamic quantities, and which are expected to be 
universal in
certain circumstances (see below). Early work on
probability distributions of correlation functions concentrated, as
numerical applications were concerned, on strictly one-dimensional
systems~\cite{dh,derrida,crisanti,ranmat}. The behaviour, under
renormalization group transformations, of the distribution
of {\it e.g.} conductivities in percolation-resistor networks~\cite{stw}, or 
interactions in spin glasses~\cite{yst} has been studied as well.  
 More recently, the probability
distributions of bulk quantities
such as energy, magnetisation, specific heat and susceptibility 
of disordered Ashkin-Teller models have
been investigated in two dimensions\cite{wd95} by Monte-Carlo simulations.
Bond distribution functions in one-dimensional quantum spin systems have been
revisited very recently\cite{hyman}.

Here we deal directly with spin-spin correlation functions on
finite-width strips of two-dimensional disordered
Ising systems. The basic motivation for using this geometry is
the fact that strip calculations, in conjunction with finite-size scaling
concepts~\cite{fisher,fs1} are among the most accurate
techniques to extract critical points and exponents for non-random
low-dimensional systems~\cite{nig82,fs2}.
The rate of decay of correlation functions
determines correlation lengths along the strip.
These latter are, in
turn, a key piece of Nightingale's phenomenological 
renormalisation scheme~\cite{nig82,fs2}, 
and have been given further relevance via
the connection with critical exponents provided by conformal invariance
concepts~\cite{cardy}.
Early extensions of strip scaling to 
random systems~\cite{early}
have since been pursued further~\cite{dQ92,sldq,sbl} and put into a broader
perspective. Though this has been done with the help of ideas arising from the study of probability distributions~\cite{dh,derrida,crisanti,ranmat},
the behaviour of the  probability distributions
themselves has not been closely investigated in strip geometries.
In particular, their
evolution toward the two-dimensional system's
form as strip width increases has not been analysed.  

We consider a two-dimensional random-bond Ising model on a square lattice
with a binary distribution of ferromagnetic interaction strengths,
each occurring with equal probability:
\begin{equation}
 P(J_{ij})= {1 \over 2} ( \delta (J_{ij} -J_0) +  \delta (J_{ij} -rJ_0) ) 
,\ 0 \leq r \leq 1\ \ .
\label{eq:1}
\end{equation}
\noindent For this case, the transition temperature $\beta_c = 1/k_B T_c$
is exactly known from duality\cite{fisch,kinzel}, 
\begin{equation}
\sinh (2\beta_{c} J_{0})\sinh (2\beta_{c}r J_{0}) = 1 \ \ .
\label{eq:2}
\end{equation}
We have studied strips  of width $L \leq 13$ sites, with periodic boundary conditions,
and length $N = 10^6$ sites. Throughout this work we
fix $r=1/4$ and $T = T_c\, (1/4)$ as given by Eq.\ (\ref{eq:2}). Numerically,
$T_c\, (1/4)/J_0 = 1.239 \dots$  (to be compared 
with  $T_c\, (1)/J_0 = 2.269 \dots$~). Using this value of $r$ ensures that disorder effects are rather strong, while at the same time one keeps
a safe distance from the percolation regime at $r =0$ (near which
crossover to geometry-dominated behaviour is expected to complicate the
picture); this choice also coincides with that used in several recent
Monte-Carlo simulations~\cite{sst,talapov},
thus comparison (when appropriate) is made easier. The choice of  $T = T_c$
is important, as it is here that the probability distributions are expected
to have a non-trivial universal form;
furthermore, the extensive literature on critical correlations for pure 
systems, both making  
explicit connection to conformal invariance ideas~\cite{cardy}
and previous to that (see {\it e.g.} Ref. ~\onlinecite{wu76} and references
therein), is an important 
reference frame against which to set our results. 

In what follows, we first illustrate the role played by intrinsic fluctuations
in the probability distribution of correlation functions, and show that even
though these do not die away for large samples, the sample-to-sample
fluctuations of averaged values do go down as sample size increases. Next we
 compare our results for averaged critical
correlations with those for a pure system, in order to check on a
recent proposal arising from Monte-Carlo data~\cite{talapov} which implies
equality, within error bars, of the corresponding quantities. 
We then go on to identify the key features of the shape of distributions, and
investigate their
variation with distance $R$ and strip width $L$. A simplified scaling theory
is formulated, which provides the essential aspects of the $R$-- and $L$--
dependence. Numerical data for the probability distributions of
correlation functions bear out the main predictions of scaling theory,
in particular the role played by the combination $R/L$ as an appropriate
scaling variable.

\section{Intrinsic fluctuations and averages}

We calculate the spin-spin correlation function 
$G\, (R)\, \equiv \, <\sigma_{0}^1 \sigma_{R}^1>$,
between  spins on the same row (say, row 1), and $R$ columns apart,
of strips with periodic boundary conditions along the vertical direction.  
This is done following the lines of Section 1.4 of Ref. \onlinecite{fs2},
with standard adaptations for an inhomogeneous system\cite{sldq}. 
At each iteration of the transfer matrix from one column to the next,
the respective vertical and horizontal bonds between first-neighbour spins
are drawn from the bond probability distribution, Eq.\ (\ref{eq:1}).
By shifting the origin along the strip and accumulating the corresponding
results, one then obtains averages of the correlation function (or of
any function $F$ of it,  such as its logarithm, which will be of particular
importance in what follows), to be denoted by $\langle G\rangle$
(or $\langle F\left( G \right) \rangle$ ), the $R$-- dependence
being implicitly understood. With strips of length $N = 10^6$
sites, we are able to produce $10^4$--$10^5$ independent estimates  of
$G$ for $7 \leq R \leq 100$ which is the range of distances to concern
us here.

 Normalized histograms of occurrence of the allowed values of
the correlation function 
(or, rather more frequent below, of its logarithm) are
produced by dividing a convenient interval of variation of $G$
(or $\ln G$) into $10^3$ equal-width bins,
and assigning each particular realization to the appropriate bin.
For $\ln G$ the interval ranging from $\ln 10^{-7}$ to zero has proved
generally adequate, except for $R=100$ where the lower limit was pushed 
down to $\ln 10^{-11}$.

For strictly one-dimensional disordered systems ({\it i. e.}, chains) the
average free energy (related to the largest Lyapunov exponent) has a normal distribution~\cite{dh,derrida,ranmat}, as befits a sum of random
variables. Thus the fluctuations shrink with sample size (strip length) $N$,
and relative errors  must die out as $1/\sqrt{N}$.
Correlation functions, on the other hand, are $products$ of random variables,
thus their distribution tends
to a $log$-normal form as $R \to \infty$~\cite{dh}, that is,
the probability distribution of $ \ln G  $ approaches
a Gaussian. However, the analysis of correlation functions turns out
to be more complex than that of the free energy, even on chains; 
a primary reason for this is that while the latter quantity is self-averaging 
in the sense defined above, the former is not: the usual Brout 
argument~\cite{brout} cannot be applied, as explained {\it e. g.} in 
Ref.~\onlinecite{derrida}. The width of the
probability distribution of correlation functions is then expected to be
a permanent feature, which will not vanish (at least trivially) with
increasing sample size.

We have found that on finite-width strips the width of the distribution tends
to stay essentially constant as $N$ varies. A graphic illustration is
provided in 
Fig. \ref{fig:1} where the horizontal variable is $\ln G $,
which turns out to be convenient for most purposes (see below).
Increasing $N$ simply smooths out the histogram;
averages such as  $\langle G \rangle$ or  
$\langle \ln G \rangle$ hardly move, the same being true
of the width. This is easier to notice by comparing the Gaussians fitted to peak
at $\langle \ln G \rangle$ and with width given by the root-mean-square
deviation
$ \Delta (\ln G) \equiv \{ \langle \left[ \ln G  - \langle \ln  G \rangle \right]^2\rangle \}^{1/2}$ .
  
Though neither $ \Delta (\ln G)$ or $ \Delta  G$  vanishes, it is
still possible to extract valuable information from averaged values, the
dispersion of which among independent samples (to be denoted respectively 
$\Delta \langle \ln G \rangle$ or
$\Delta \langle G \rangle$) $does$ shrink with increasing sample size. Fig. \ref{fig:2}
shows the typical dependence of relative fluctuations,
$ \Delta  G / \langle G \rangle $ and
$\Delta \langle G \rangle /  \langle G \rangle $, with strip length
$N$. Varying the number $n$ of distinct
(that is, $N$--lattice spacings long) samples within
a reasonable interval, say $n = 5$--$50$, changes
$\Delta \langle G \rangle$ only slightly, consistent with  a $1/\sqrt{n}$--dependence to be inferred from standard arguments. From an
investigation of $\Delta \langle G \rangle /  \langle G \rangle $
for distances in the
range $R =7$--$50$ and strip widths up to $L = 7$, it turns out that both the 
order of magnitude and $N$-dependence (roughly $1/\sqrt{N}$)
depicted in Fig. \ref{fig:2}  are typical.  The behaviour of 
$\Delta \langle \ln G \rangle /  \langle \ln G \rangle $ is
entirely similar. We can thus predict (see Fig. \ref{fig:2})
that the fluctuations 
$\Delta \langle \ln G \rangle /  \langle \ln G \rangle $ 
 and $\Delta \langle G \rangle /  \langle G \rangle $
will be of order $1\%$ or just under that for strips of length $N = 10^6$.
This will be
enough for our purposes here. Similar considerations have been used 
elsewhere in strip studies~\cite{sbl}, and seem to have been followed also in
Monte-Carlo calculations of correlation functions in finite ($ L \times L$)
systems~\cite{talapov}.

\section{Comparison with pure-system critical correlations}
It has been found in Monte-Carlo simulations~\cite{talapov} that the average correlation function at criticality of a random-bond Ising system
is numerically very close to that for a pure system at its own critical point.
Below we check on the corresponding quantities for the strip geometry.

The spin-spin correlation function for the pure Ising model at $T = T_c$
on a strip of width $L$ is known from conformal invariance~\cite{cardy} to 
vary, for large $R$, $L$ as:
\begin{equation}
<\sigma_{0}^1 \sigma_{R}^1>\ \sim \left( {\pi/L \over \sinh (\pi R/L)}\right)^{\eta} \ \ ,\ \  \eta =1/4,
\label{eq:3}
\end{equation}
\noindent for spins along the same row as is the case here.
The proportionality factor can be obtained from the exact square--lattice 
($L \to \infty$) result~\cite{wu76}, 
$<\sigma_{0}^1 \sigma_{R}^1> = 0.70338/R^{1/4} $. In  Fig. \ref{fig:3}
we show, for $R=7$ and 20, data for $\langle G \rangle$ and 
$\exp \langle \ln G \rangle$ together with  a
continuous curve for the pure system. The latter passes through numerically
calculated points for $L \leq 15$ ( Eq.\ (\ref{eq:3}) is in error by
one part in $10^4$ for $L=15$, $R=7$ and less than that for $R=20$)
and follows  Eq.\ (\ref{eq:3}) for larger $L$. Using $1/L^2$
for the horizontal axis guarantees that the pure-system curve approaches
the vertical axis linearly. However, it still shows high curvature
even for the largest values of $L$ attainable in our random-system
calculations. This warns us to refrain from extrapolating our data for
$L \to \infty$. Even so, we can learn from finite-width results that
$\langle G \rangle$ behaves very closely to its pure-system counterpart. This is
in line with previous findings~\cite{sldq} according to which in a random system
the correlation length $\xi$
to be used in the exponent--amplitude relation of conformal invariance,
$\xi = L/\pi \eta$~\cite{cardy}, is that obtained from the average decay of
$\langle G \rangle$ against $R$. Thus one
gets a picture in which $\eta = 1/4$ as for the pure system~\cite{sldq}, consistent with
$\gamma/\nu =7/4$ obtained {\it e. g.} from strip calculations of the average 
susceptibility for the random system~\cite{sbl},
and the scaling relation $\gamma/\nu = 2 - \eta$.

Of course the present
result reaches further, as one could conceive of a scenario where the
correlations would differ in the pure and random systems, but decay 
asymptotically as $R \to \infty$ with the same rate (thus giving the 
same $\xi$).  In fact, the decay of $\exp \langle \ln G \rangle$ against $R$
is not too dissimilar to that of $\langle G \rangle$ for moderate disorder,
and for the finite values of $L$ within reach of calculation. Only a
systematic study of extrapolation trends as $L \to \infty$, covering different
degrees of disorder, shows how the respective correlation lengths are
essentially distinct~\cite{sldq}. The physical origin of this lies in that,
on account of the properties of the probability distribution of $G$ (to be
seen in detail below),
the {\it most probable} value $\exp \langle \ln G \rangle$ does not 
coincide with the {\it average} one, $\langle G \rangle$~\cite{crisanti,ranmat}.
Accordingly, it has been shown by field-theoretic arguments~\cite{ludwig}
and supported by numerical work~\cite{sldq} that the most probable, or
typical, correlation function decays as
\begin{equation}
 \exp \langle \ln G \rangle \sim R^{-1/4} (\ln R)^{-1/8} \ \ \ ,
\label{eq:ludwig}
\end{equation}
\noindent while the logarithmic corrections are washed away upon 
averaging for  $\langle G \rangle$, resulting in a purely
algebraic decay with $\eta = 1/4$.

Quantitative analysis of the results displayed in Fig. \ref{fig:3} shows that ,
considering the central estimates $\langle G \rangle$ the
ratio $ Q \equiv \langle G\, (R,L,r,T_c(r))\rangle / G\, (R,L,1,T_c(1)) $ is, in all cases, within $1.01 - 1.03$.
 With estimated error bars of order $1\%$
as explained in the previous Section, 
 a very small amount of overshooting seems to persist which does not
follow a definite trend against $R/L$ (see Fig. \ref{fig:4}).
 Monte-Carlo data show the corresponding ratio
approaching unity from below as lattice size $L$ increases~\cite{talapov},
in the region $R/L \ll 1$. We cannot go far into that region, as the maximum
strip widths within reach are not much larger than 10, and randomness
effects are significantly distorted for small $R \lesssim 5$. 

Though we are not able at this point
to advance a quantitative argument, it is plausible that
 finite-size effects manifest themselves differently in strip and
square geometries; taken 
together with those of Ref. \onlinecite{talapov}, we interpret our data
as evidence in favour of pure-- and random--system critical correlations
being in fact equal, at least  for $R$, $L \gg 1$ and $R/L \ll 1$.

In the next two Sections, we exploit the features of the probability
distribution of $G$, and show that the variable $R/L$ is indeed 
at least approximately the convenient one to describe several 
relevant aspects of the problem.   

\section{Probability distributions: scaling theory}
\subsection{Relevant parameters}
Our starting point is the result that, in one dimension and
for large $R$ the probability distribution of $G$ must be
log-normal~\cite{dh,derrida,crisanti,ranmat}. The same is expected to hold 
on strips provided that $R/L \gg 1$. We seek for
deviations from Gaussian behaviour as one moves away from this limit.

In Fig. \ref{fig:5}
we show normalized histograms of occurrence of $\ln G$ for fixed $L=5$
and $R=7$, 20 and 50 . Though to zeroth-order one could say that all plots look
similar to Gaussians, the semblance is reduced as $R$ decreases.

A quantitative measure of departure from Gaussian behaviour is the
(dimensionless) {\it skewness} $S$, defined as~\cite{numrec}: 
\begin{equation}
 S \equiv \left({\langle x - \langle x \rangle \rangle \over \sigma}\right)^3
\label{eq:4}
\end{equation}
\noindent for a distribution with mean $ \langle x \rangle $ and
dispersion $ \sigma $. Of course, for a finite number of realizations
of a given probability distribution $S$ itself will be subject to 
fluctuations. In what follows we shall always quote $S$ with two significant
digits, which will allow us to discern trends while staying reasonably
within reliable margins. For $L=5$ and  $R=7$, 20, 50 and 100 (the latter
not shown in Fig. \ref{fig:5}) one has respectively $S= -0.67$, $-0.41$,
$-0.24$ and $-0.19$. We shall analyse the $R-$ and $L-$ dependence of $S$
below; for the moment note that it approaches zero with increasing $R$,
as expected, and is always $negative$. This is partly because of the 
many-channel, incipiently two-dimensional character of correlations on the
strip: qualitatively, if there is at least one path of
 ``strong'' bonds between two
spins, their correlation is significant, and is not much enhanced if there
are more strong--bond paths, thus the peak at large $\ln G$ with
an abrupt cutoff above the maximum; on the other hand, configurations without
any strong--bond paths at all are possible but with low probability, giving
rise to the ``tail'' of very low  $\ln G$ values. This argument explains
the general trend towards increasing $|S|$ for $R/L < 1$ as well (see below).
An extreme example of this trend in shown in  Fig. \ref{fig:6} for $L=13$, $R=7$
where $S=-0.85$. 
The negative skewness effect
depends also on the variable against which histograms are plotted: here we use
 $\ln G$ because our goal is to check on departures from a log-normal
distribution, thus it is the skewness of this plot which matters in the context.
For instance, a plot of the same distribution against $\tanh^{-1} G$ would
have positive skewness.  

We choose to characterize the distribution of $\ln G$ by three
quantities, namely mean ($\langle \ln G \rangle$), dispersion, or
root-mean-square deviation ($ \Delta (\ln G)$ ) and skewness ($S$). In other
words, we assume that the  probability distribution of correlation functions
is satisfactorily described by perturbative corrections to a log-normal form.  
As shown below, this works well in the present case of ferromagnetic disorder.
If frustration effects are present, such as
{\it e. g.} in random-field~\cite{ourrf} or spin-glass systems,
it may be necessary to take recourse to additional parameters, or even 
to adopt a different perspective. We shall not deal with this matter in the
present work. 

\subsection{Scaling Theory}

The aspects just described are consistent with a scaling description
given in this subsection. It leads to further specific predictions regarding
the form of distribution functions and their dependences on the variables
$R$ and $L$, which will be discussed subsequently in the light of the
numerical results.

The approach combines two main features: $(i)$\, the appearance at large scales
$L$, $R \gg 1$ of universal aspects related to fixed point Hamiltonian and
probability distributions of the disordered two-dimensional ($2d$) Ising model;
and
$(ii)$\, the crossover for $R>L$ to width--limited behaviour characteristic
of the one-dimensional ($1d$) version of the large scale universal properties.

The procedure is to first scale $n$ times by (length) rescaling factor $b$,
where $b^n=L$, which takes the system to an equivalent linear chain. This
step involves the scaling of joint probability distributions for the
appropriate variables. If one starts from the critical condition of the
random $2d$ Ising system irrelevant variables scale away, and one approaches
asymptotically the fixed point Hamiltonian and distribution describing the
universality class containing the $2d$ random Ising system. This involves a
minimal set of relevant random variables$\{ t_i^{(n)}\}$ (after $n$
scalings) and their universal probability distributions. The correlation
function scales as follows:
\begin{equation}
G_L\,(R,\{t_i\}) = b^{-\eta}\, G_{L/b}\,(R/b,\{t_i^{(1)}\}) = \dots  
=L^{-\eta}\, G_1\,(R/L,\{t_i^{(n)}\})\ \ \ (n=\ln L /\ln b )\ .
\label{eq:cfsc}
\end{equation}
We now have an equivalent $1d$ system with Hamiltonian close to the fixed point
Hamiltonian of the disordered $2d$ system. Since the correlations in a $1d$
system are transmitted through each intermediate space point, a factorisation
of $G_1\,(R/L,\{t_i^{(n)}\})$ is suggested if $R/L>1$. This factorisation is
into $R/L$ factors corresponding to successive $L \times L$ blocks of the
original system ({\it i.e.} to single bonds of the renormalised $1d$
system), labelled by $s=1$, $\dots$, $R/L$;  then Eq.\ (\ref{eq:cfsc}) becomes
\begin{equation}
G_L\,(R,\{t_i\}) = L^{-\eta}\,\prod_{s=1}^{R/L}G_1\,(1,\{t_i^{(n)}\}_s)\ .
\label{eq:cfsc2}
\end{equation}
\noindent Here, $\{t_i^{(n)}\}_s$ are the renormalised random variables for
the block labelled by $s$. It will be assumed later that these are largely
uncorrelated from one block $s$ to another, since the blocks were 
initially non-overlapping. This, and the other approximations leading to the
approximate form  Eq.\ (\ref{eq:cfsc2}) will be tested by later comparison to
the numerical results. Eq.\ (\ref{eq:cfsc2}) generalises a result for the pure
case. There, for $R/L$ large, $G_1\,(R/L,\dots t^{\ast} \dots) \propto
(\lambda_2/\lambda_1)^{R/L}$, where $\lambda_2/\lambda_1$ is a universal ratio
of eigenvalues of the transfer matrix of the universal fixed point
Hamiltonian of the pure $2d$ Ising class. Then
\begin{equation}
 G_L\,(R,t^{\ast}) \propto L^{-\eta}\, \exp [-R/\xi_L]\ ,
\label{eq:cfsc3}
\end{equation}
\noindent where $\xi_L^{-1} = {1 \over L}\ln (\lambda_1/\lambda_2)$
($=\pi \eta/L$~\cite{cardy}). It is perhaps interesting that the simplest $b=2$
Migdal-Kadanoff real-space renormalization group transformation gives 
$G_1\,(R/L,t^{\ast}) = (t^{\ast})^{R/L}$, $t^{\ast} = 0.544$, hence it gives
$\eta \sim 0.19$. This unsatisfactory representation of the universal $\eta$
in terms of a non-universal $t^{\ast}$ is due to not having allowed the
Hamiltonian to adopt its universal form.

We now return to the disordered case and consider the development of the
probability distributions during the rescalings leading to Eq.\ (\ref{eq:cfsc}).
To allow for correlations it is necessary to consider the probability
distribution $P_t\{t_i\}$ for the whole set of $t_i$'s.
This is labelled by a parameter $t$ setting the scale for all the $t_i$'s.
 The scaling of the
distribution is given by a mapping $W_b$:
\begin{equation}
P_{t^{(l+1)}}^{(l+1)}\,\{t_i\} = W_b\left\{P_{t^{(l)}}^{(l)}\,\{t_i\}\right\} =
 W_b^l\left\{P_{t^{(0)}}^{(0)}\,\{t_i\}\right\}\ ,
\label{eq:cfsc4}
\end{equation}
\noindent where $l$, $l+1$ label two successive steps, and $W_b^l$ denotes
$l$ iterations of the map. The parameter $t$ also scales according to a
renormalization group transformation characteristic of the $2d$ random Ising 
model. At the fixed point $t^{\ast}$ of that transformation, after
$n =\ln L /\ln b$ scalings with $n$ large,  $W_b^n$ will have produced a
distribution $P_{t^{\ast}}^{(n)}$ close to the universal invariant
distribution $P_{t^{\ast}}^{\ast}$ of the random $2d$ Ising model, which
satisfies the fixed point equation
\begin{equation}
P_{t^{\ast}}^{\ast}\,\{t_i\} = W_b\left\{P_{t^{\ast}}^{\ast}\,\{t_i\}\right\}\
 . 
\label{eq:cfsc5}
\end{equation}
\noindent Then, employing Eq.\ (\ref{eq:cfsc2}) and taking logarithms to
obtain a sum of random variables on the right-hand side we find
that at $t^{\ast}$, after $n$ scalings, the probability distribution
${\cal P}(\alpha)$ for $\ln G_L$ ({\it i.e.}, for the probability that
$\ln G_L\, (R, \{t_i\})$ takes the value $\alpha$) is given by:
\begin{equation}
\int d\alpha\, \exp(\beta\alpha){\cal P}(\alpha) =
\int\left(\prod dt_i\right)\, P_{t^{\ast}}^{(n)}\,\{t_i\}\exp(-\beta\eta\ln L)
\prod_{s=1}^{R/L}\,\exp(\beta\ln G_1(1,\{t_i\}_s) \ . 
\label{eq:cfsc6}
\end{equation}
If $n$ is large, as well as $P_{t^{\ast}}^{(n)} \sim P_{t^{\ast}}^{\ast}$,
we have $G_1(1,\{t_i\}_s)$ close to a universal function characteristic of
the random $2d$ Ising class, since the renormalised bond regions labelled by
each $s$ are each a composition of many original bonds, at $t^{\ast}$.

To the extent that the probability distribution $P_{t^{\ast}}^{(n)}$
factorises into parts $q_{t^{\ast}}^{(n)}\,\{t_i\}_s$ corresponding to the
different $s$'s (to be tested) the result  Eq.\ (\ref{eq:cfsc6}) reduces to a
form corresponding to the probability of a sum of random variables
($\ln G_1$):
\begin{equation}
\int d\alpha\, \exp(\beta\alpha){\cal P}(\alpha) = \exp(-\beta\eta\ln L)
\left(I^{(n)}(\beta)\right)^{R/L} \ , 
\label{eq:cfsc7}
\end{equation}
\noindent with
\begin{equation}
 I^{(n)}(\beta) = 
\int\left(\prod dt_i\right)_s\, q_{t^{\ast}}^{(n)}\,\{t_i\}_s\exp(\beta\ln G_1(1,\{t_i\}_s)) \ ,
\label{eq:cfsc8}
\end{equation}
\noindent where the subscript $s$ indicates that all $t_i$'s  are in the
region corresponding to a given $s$.

We now explore the consequences of the above results, and in particular the
expected universality of $q_{t^{\ast}}^{(n)}$, $G_1(1,\{t_i\}_s)$ for
$L$ ($=b^n$) large, for the distribution ${\cal P}(\alpha)$ for  
$\ln G_L(R,\{t_i\}_s)$~.

Eq.\ (\ref{eq:cfsc7}) shows that  ${\cal P}(\alpha - \eta \ln L)$ 
corresponds to
the probability
distribution for a sum of $R/L$ independent random variables
($\ln G_1(1,\{t_i\}_s)$), and has the consequence that,
if $R/L$ is large, ${\cal P}(\alpha)$ approaches a
{\em Gaussian} distribution with mean, width and skewness given by:
\begin{equation}
\langle \alpha\rangle = \left({R \over L}\right) m -\eta \ln L
\label{eq:cfsc9}
\end{equation}
\begin{equation}
\langle (\alpha - \langle\alpha\rangle)^2\rangle^{1/2} = \left({R \over
L}\right)^{1/2}w
\label{eq:cfsc10}
\end{equation}
\begin{equation}
\langle (\alpha - \langle\alpha\rangle)^3\rangle/\langle (\alpha - \langle\alpha\rangle)^2\rangle^{3/2} = \left({R \over L}\right)^{-1/2}s \ . 
\label{eq:cfsc11}
\end{equation}
\noindent The quantities $m$, $w$, $s$ are characteristics of
$I^{(n)}(\beta)$,
which is related as follows to the (non-Gaussian) distribution function
$Q^{(n)}(\alpha)$ for the logarithm of the nearest-neighbour
correlation function of the
($n$-times rescaled) disordered $2d$ Ising system:
\begin{equation}
I^{(n)}(\beta) =\int d\alpha\, \exp(\beta\alpha)\, Q_1^{(n)}(\alpha)\ ,
\label{eq:cfsc12}
\end{equation}
\noindent with
\begin{equation}
Q_1^{(n)}(\alpha) =\int\left(\prod dt_i\right)_s\, q_{t^{\ast}}^{(n)}\,\{t_i\}_s \,\delta(\alpha - \ln G_1(1,\{t_i\}_s))\ .
\label{eq:cfsc13}
\end{equation}
For $L=b^n$ large, all these quantities and therefore $m$, $w$, $s$ will
become universal (characteristic of the $2d$ random Ising class).

The universal character at large $L$ of the distribution ${\cal P}(\alpha)$
for $\ln G_L\,(R)$ at $t^{\ast}$, its Gaussian form at large $R/L$
and the results (\ref{eq:cfsc9})--(\ref{eq:cfsc11}) and their interpretation
above are the main conclusions of this Section. Note $(a)$ that these
conclusions include the result of Ref.~\onlinecite{dh} for the $1d$ case and
$(b)$ that the analogue of $m$ for the pure case is the universal constant
$-\pi\eta$ (while $w$, $s$ are zero).

The essential points of the general discussion given above can be explicitly
illustrated in the simple scenario provided by the renormalization group
approach (blocking/decimation), allowing for just the random variables
$\{t_i \equiv \tanh \beta J_i\}$. This forces the Hamiltonian to remain
of Ising form (so the reservation expressed under
Eq.\ (\ref{eq:cfsc3}) applies). We allow for spin rescaling, and for the
distribution function $g(t_i)$ for each $t_i$ to develop towards fixed point
universal form under the scaling~\cite{stw,yst}, but we ignore correlations.
Then  $G_1(1,\{t_i\})$ is just the renormalized variable $t_i^{(n)}$. The
scaling of $t_i$ will be of the form
\begin{equation}
t_i^{(l+1)} = R_b\{t_i^{(l)}\}
\label{eq:cfsc14}
\end{equation}
\noindent where the right-hand side is a function of $N_b$ variables 
$t_i^{(l)}$ comprising the block. Then the distribution $g^{(l)}(t_i^{(l)})$
scales according to the following simplified version of Eq.\ (\ref{eq:cfsc4}):
\begin{equation}
g_{t^{(l+1)}}^{(l+1)}(t^{\prime}) = \int\left(\prod_{i=1}^{N_b} dt_i\,  g_{t^{(l)}}^{(l)}(t_i)\right) \delta(t^{\prime}-R_b\{t_i^{(l)}\})\ ,
\label{eq:cfsc15}
\end{equation}
\noindent and $t^{(l)} \to t^{(l+1)}$ is the resulting change of scale of
the distribution $g$. No scale change occurs if the initial distribution
is set at the critical value of $t^{(0)}$. Then, for large $l$,
Eq.\ (\ref{eq:cfsc15}) gives the asymptotic approach to the universal fixed
point distribution, from which $m$, $w$, $s$ can be obtained. For an adequate
description of this sort, the transformation Eq.\ (\ref{eq:cfsc14}) should
give the proper zero value of the exponent $\alpha$ for the pure case: via the
Harris criterion this makes the disorder marginally relevant and ensures that
the width $w$ does not scale away. It requires $N_b = \lambda_b^2$, where
$\lambda_b = b^{1/\nu}$ is the eigenvalue of the pure version of 
Eq.\ (\ref{eq:cfsc14}), linearized about its fixed point. Procedures of this sort
give $(i)$ $m$ not very different from its pure value; $(ii)$ negative
skewness $s$. 

\section{Numerical Results and contact with the scaling theory}

We begin by recalling that the results exhibited in
 Fig. \ref{fig:2} point out the importance of
intrinsic widths in the critical random system ({\it i.e.} at $t^{\ast}$).
This is a central feature of the scaling theory, through the appearance of
universal distributions. Secondly, we recall that Fig. \ref{fig:5}
provides evidence for Gaussian distributions for $\ln G$ at large $R$, with
narrowing relative widths as $R$ increases. This is again a prediction of
the scaling theory (see Eq.\ (\ref{eq:cfsc9}) and the discussion preceding it).

The results already presented in Fig. \ref{fig:3} show that
the overall dependence of
$\exp \langle \ln G \rangle$ on $L$ for fixed $R$ approximately mimics that of
$\langle G \rangle$, apart from a proportionality factor. The latter is, in
turn, numerically very close (see Fig. \ref{fig:4}) to that of the pure system,
given to good approximation by Eq.\ (\ref{eq:3}). The proportionality factor
is, however, $R-$ dependent, as illustrated in Fig. \ref{fig:7}; this
illustrates that, though both quantities decay exponentially with $R$
as befits an
essentially one-dimensional system, their respective correlation lengths 
differ, with well-known consequences~\cite{crisanti,ranmat,sldq}.
The linear dependence of $\langle \ln G\rangle \equiv \alpha$ against $R$
is consistent with the scaling result (\ref{eq:cfsc9}).

Also, at large $R$, the slope of the measured $\alpha$ versus $R$ line is
proportional to $1/L$, in agreement with the $(R/L)m$ term dominant in 
Eq.\ (\ref{eq:cfsc9}); and the coefficient is consistent with having $m$
not far from its pure value. The comparison of numerical results for
$\langle \ln G\rangle$ and $\langle G\rangle$ with corresponding pure forms is
given in Fig. \ref{fig:8} where $G_{pure}$ is given by Eq.\ (\ref{eq:3}) above.

A further, rather stringent test of the scaling predictions is
the plot in Fig. \ref{fig:9} of numerical data for
$ \Delta (\ln G)$ against $R/L$. It can be seen that the $(R/L)^{1/2}$
dependence Eq.\ (\ref{eq:cfsc10}) is reasonably followed
in the large $R/L$ regime where it was derived; and the data collapse of
results for different $R$, $L$ suggests that $R/L$ is (as predicted) the
appropriate scaling variable in the regime of large $R$, $L$ independent of
their ratio. The curve also gives evidence of the crossover to the universal
width of the $2d$ random Ising system for $L \gtrsim R \gg 1$.

In order to further test the scaling theory, and the suggestion that
$R/L$ is the apppropriate scaling variable,  we show in
Fig. \ref{fig:10} numerical results for the skewness against  $R/L$.
Again the data collapse is satisfactory. The prediction of
$(R/L)^{-1/2}$ behaviour (Eq.\ (\ref{eq:cfsc11})) for large $R/L$ is again 
observed, and the crossover towards universal $2d$ behaviour is seen
for $L \gtrsim R \gg 1$.

\section{Conclusions}

We have studied properties of the probability distributions of
correlation functions on finite-width strips of the two-dimensional random-bond
Ising model at criticality. 
We have shown that even though intrinsic fluctuations
in the probability distribution do not die away for large samples, the sample-to-sample
fluctuations of averaged values do go down approximately
with the square root of sample size as the latter increases. 
Results thus obtained for averaged critical
correlations have been compared with those for a pure system, and we have
found  that the values of averaged correlations $\langle
G\rangle$ are very close to the corresponding pure-system ones,
consistent with recent Monte-Carlo data~\cite{talapov}. 
The key features of the shape of distributions have been identified,
and a simplified scaling theory
has been formulated, which provides the essential aspects of the $R$-- and $L$--
dependence. Numerical data for the probability distributions of
correlation functions bear out the main predictions of scaling theory,
in particular the role played by the combination $R/L$ as an appropriate
scaling variable.

We expect the approach outlined above, which consists in describing
the probability distribution of  $\ln G$
by perturbative corrections to a log-normal form
(thus characterized by only three
quantities, namely mean ($\langle \ln G \rangle$), width ($ \Delta (\ln G)$ )
and skewness ($S$))
to be appropriate in the present case of ferromagnetic disorder;
it remains to be checked whether additional parameters, or even a change of
perspective, will be necessary if frustration effects are present, 
{\it e. g.} in random-field~\cite{ourrf} or spin-glass systems. We plan to 
undertake this task as a continuation of the present work.

\acknowledgements

SLAdQ thanks the Department of Theoretical Physics
at Oxford, where most of this work was carried out, for the hospitality, and
the cooperation agreement between Academia Brasileira de Ci\^encias and
the Royal Society for funding his visit. Research of SLAdQ
is partially supported by the Brazilian agencies Minist\'erio da Ci\^encia
e Tecnologia, Conselho Nacional
de Desenvolvimento Cient\'\i fico e Tecnol\'ogico and Coordena\c c\~ao de
Aperfei\c coamento de Pessoal de Ensino Superior.

\newpage

\begin{figure}
\caption{
Normalized histograms of occurrence of $\ln G$ for $L=5$, $R=20$. $(a)$ : 
$N = 10^5$; $(b)$ : $N=10^6$. Full vertical arrows  at $\langle \ln G \rangle$;
broken vertical arrows at $\ln \langle G \rangle$ . Curves are Gaussians
fitted to mean and root-mean-square deviation of $\ln G$, as calculated from
respective realizations.  Here, and in all
subsequent figures, $r=1/4$ and $T=T_c(r)$.
}
\label{fig:1}
\end{figure}

\begin{figure}
\caption{
Relative fluctuations within sample, $ \Delta  G / \langle G \rangle $, and
between sample-averaged values, 
$\Delta \langle G \rangle /  \langle G \rangle $, against  strip
length $N$. $L=7$, $R=20$, number of samples $n =20$.
}
\label{fig:2}
\end{figure}

\begin{figure}
\caption{
Averaged correlation functions for $L=5$, 7, 9, 11, 13. $(a)$ : 
$R=7$; $(b)$ : $R=20$. Triangles:
$\langle G \rangle$; squares: $\exp \langle \ln G \rangle$. Errors
as defined in Section II. Continuous line:
correlation functions for pure Ising model on strips~\protect{\cite{cardy,wu76}}
.
}
\label{fig:3}
\end{figure}

\begin{figure}
\caption{
 Ratio $ Q \equiv \langle G\, (R,L,r,T_c(r))\rangle / G\, (R,L,1,T_c(1)) $
against $R/L$ for points of Fig. \protect{\ref{fig:3}}. Error bars represent
estimated fluctuations of order $1\%$.
} 
\label{fig:4}
\end{figure}

\begin{figure}
\caption{
Normalized histograms of occurrence of $\ln G$ for fixed $L=5$
and $R=7$, 20 and 50 . Full vertical arrows  at $\langle \ln G \rangle$;
broken vertical arrows at $\ln \langle G \rangle$ . Curves are Gaussians
fitted to mean and root-mean-square deviation of $\ln G$, as calculated from
respective realizations. 
} 
\label{fig:5}
\end{figure}

\begin{figure}
\caption{
Normalized histogram of occurrence of $\ln G$ for  $L=13$
and $R=7$. Full vertical arrow at $\langle \ln G \rangle$;
broken vertical arrow at $\ln \langle G \rangle$ . Curve is Gaussian
fitted to mean and root-mean-square deviation of $\ln G$, as calculated from
respective realization. Skewness $=-0.85$ . 
} 
\label{fig:6}
\end{figure}

\begin{figure}
\caption{
Semi-log plot of decay of $\langle \ln G \rangle$ and $\ln \langle G \rangle$ 
against distance for fixed $L=5$. Broken line gives slope as predicted by
conformal invariance ($L/\xi =\pi\eta$) with $\eta=1/4$.
} 
\label{fig:7}
\end{figure}

\begin{figure}
\caption{
 Log-log plot of ratio between averaged correlation functions and $G_{pure}$ against $R/L$, the latter as given in Fig.
\protect{\ref{fig:3}}~\protect{\cite{cardy,wu76}}. Triangles: data from
$\langle G\rangle$; squares: data from $\langle \ln G\rangle$. 
} 
\label{fig:8}
\end{figure}

\begin{figure}
\caption{
Log-log plot of  width $ \Delta (\ln G)$ against distance $R/L$~. Line
has slope 1/2 .
} 
\label{fig:9}
\end{figure}

\begin{figure}
\caption{
Log-log plot of negative skewness $-S$ against $R/L$~. Line has slope $-1/2$ .
} 
\label{fig:10}
\end{figure}

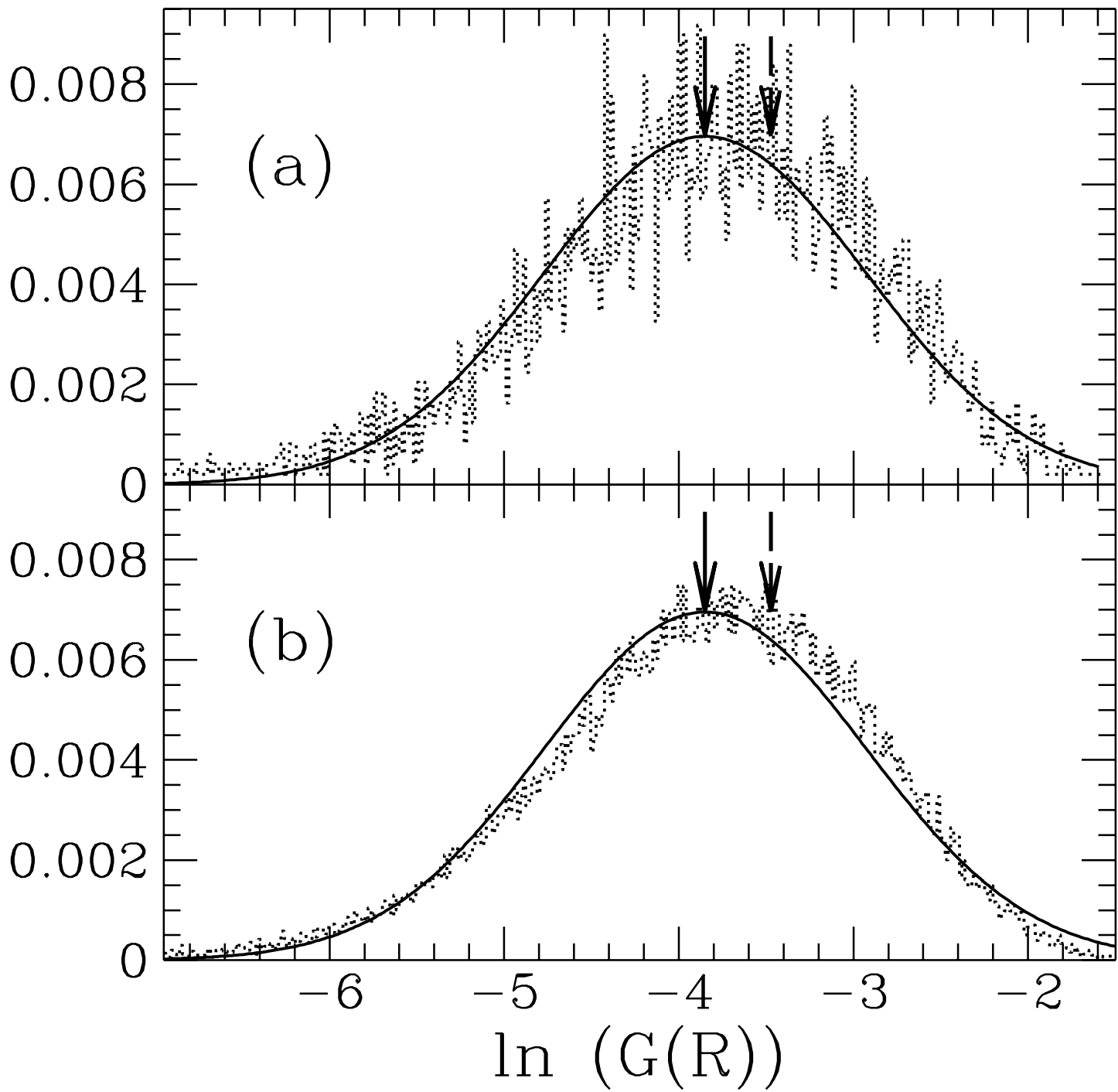
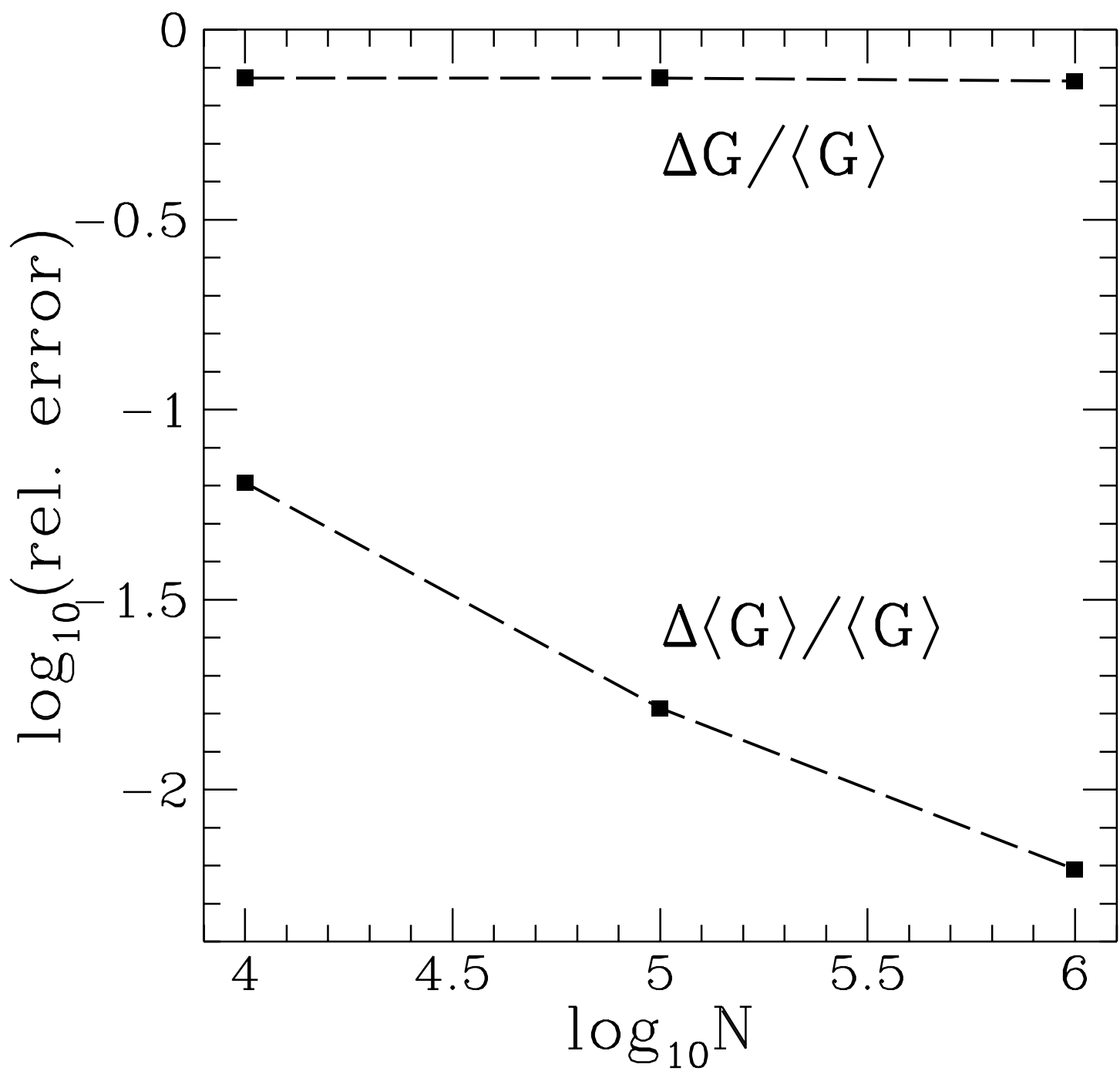
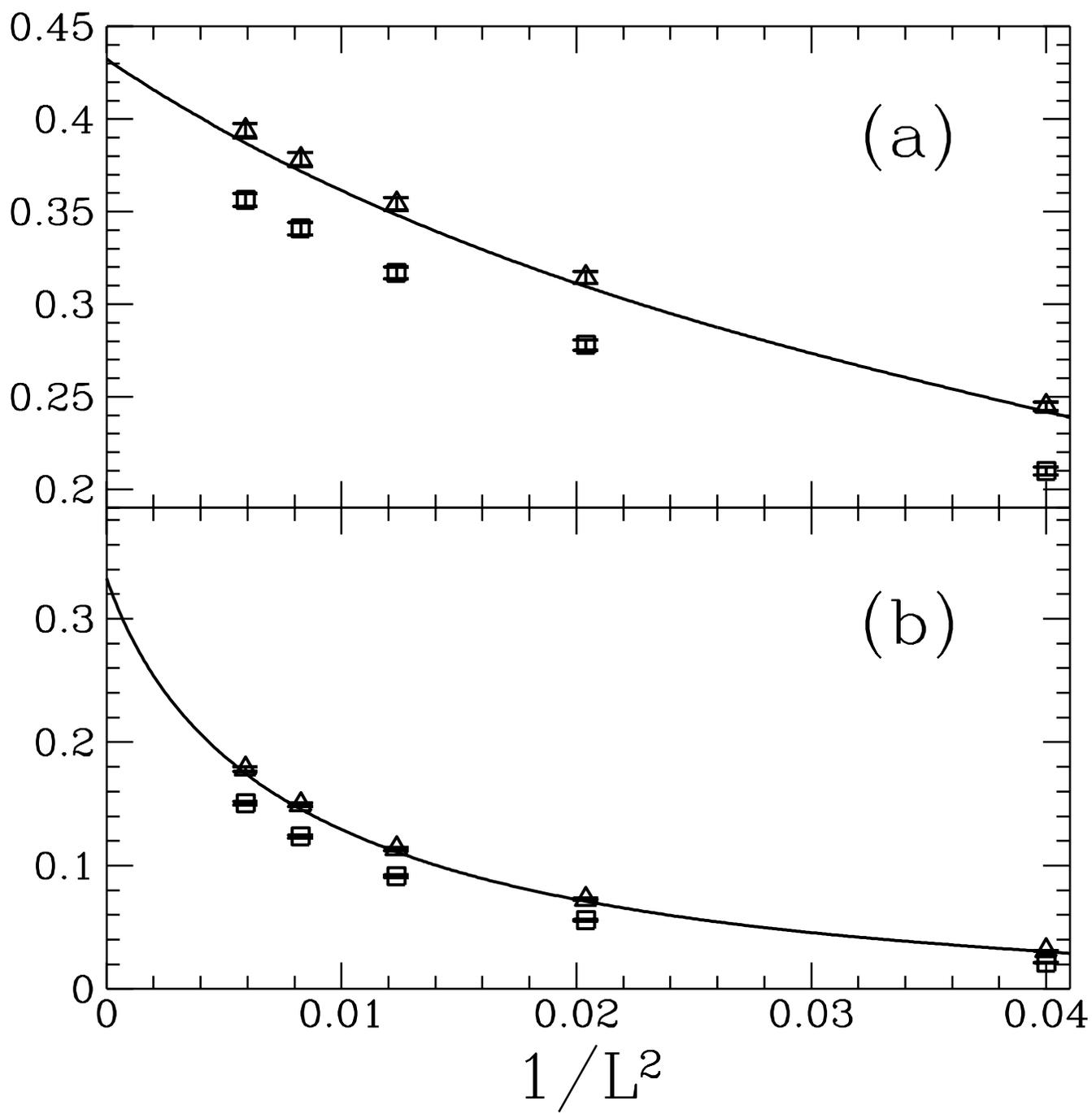
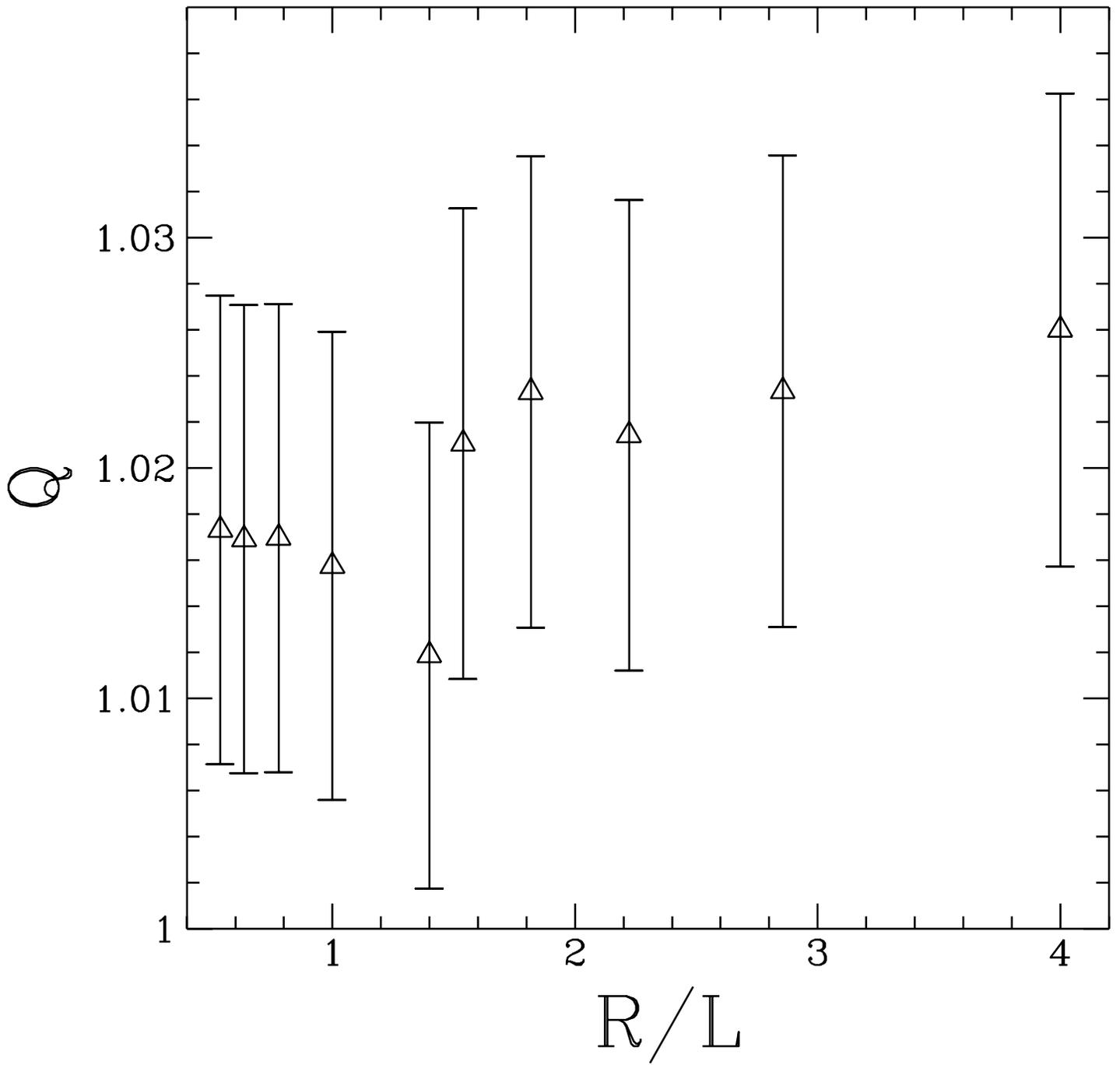
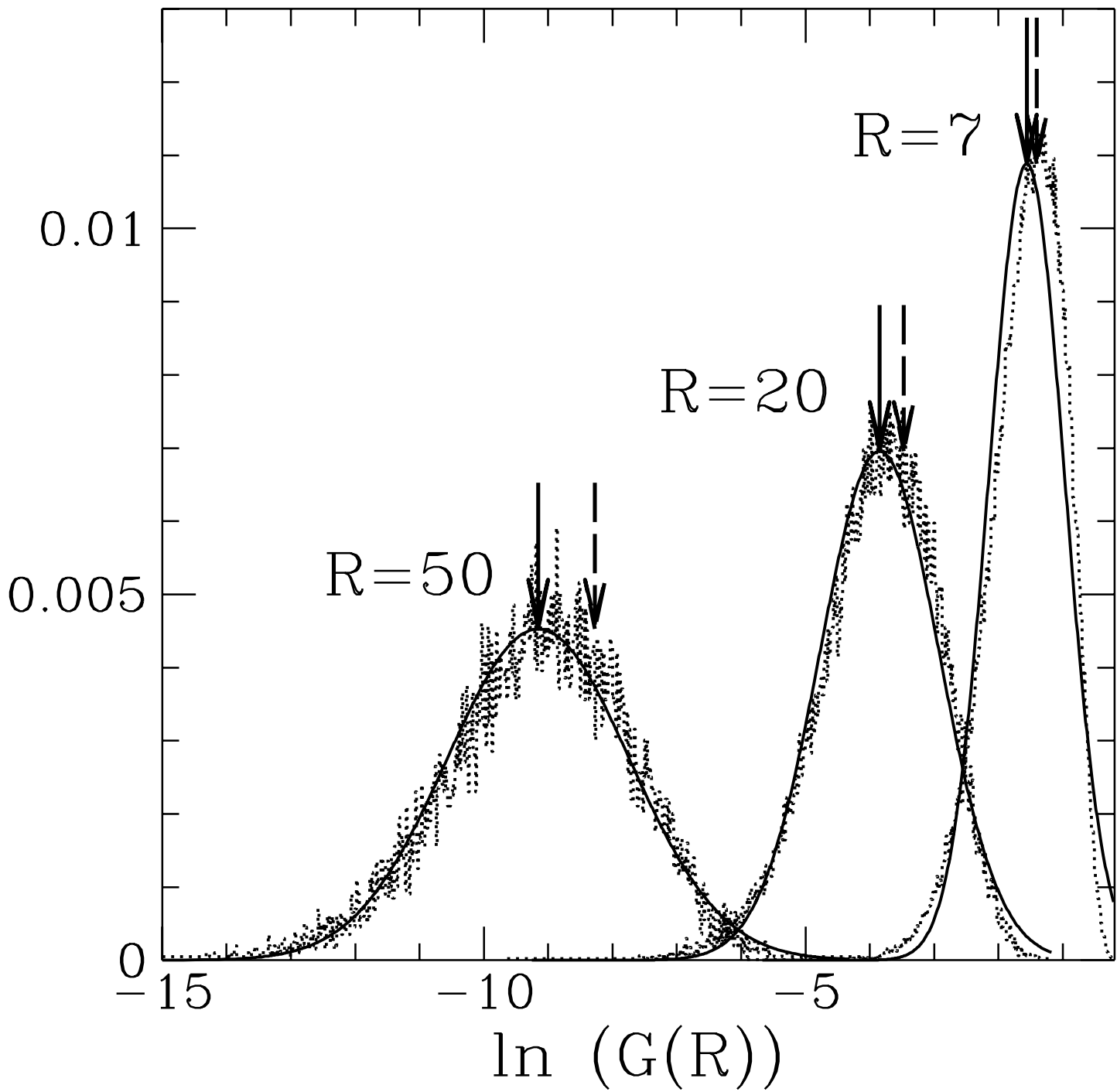
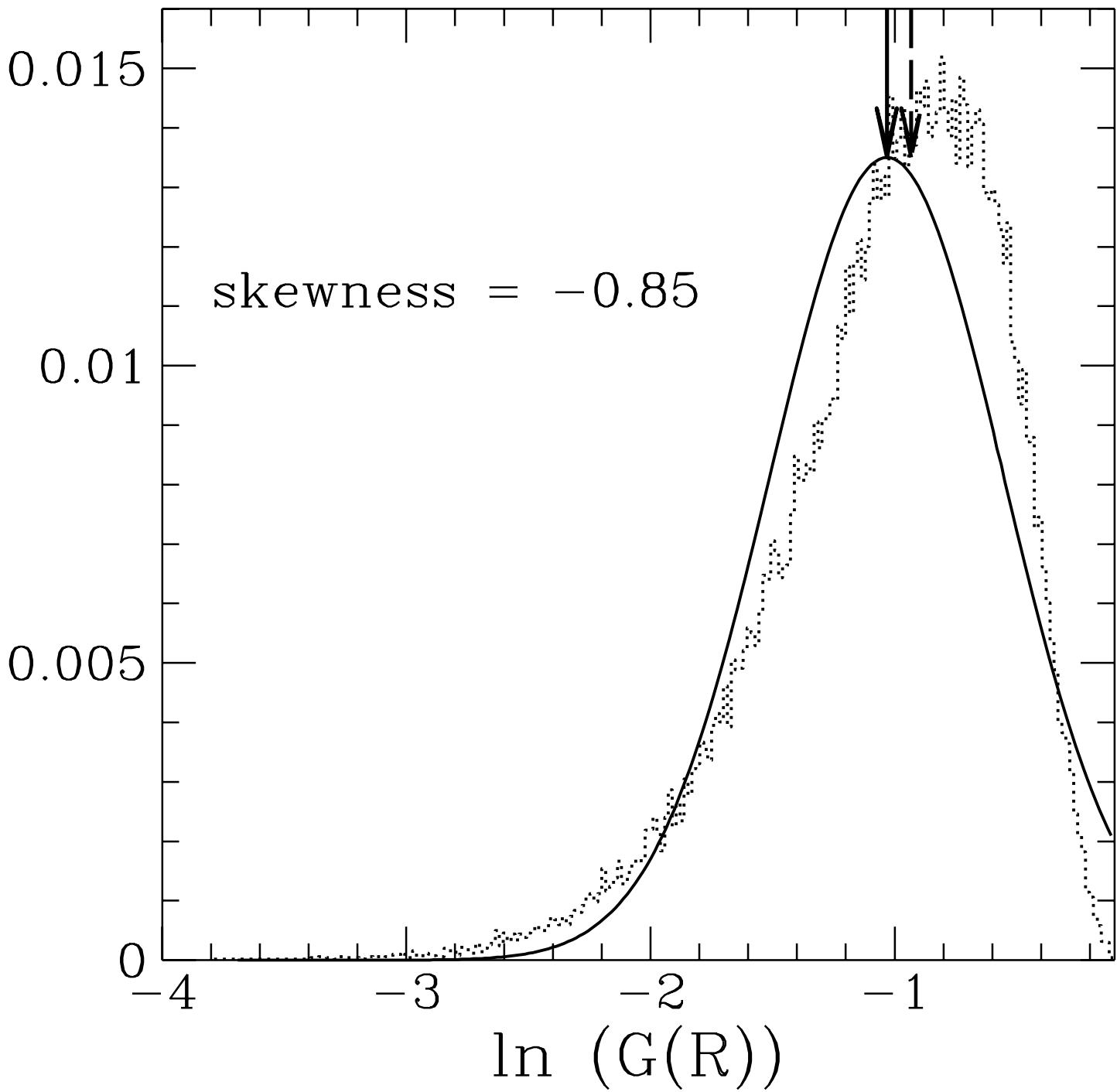
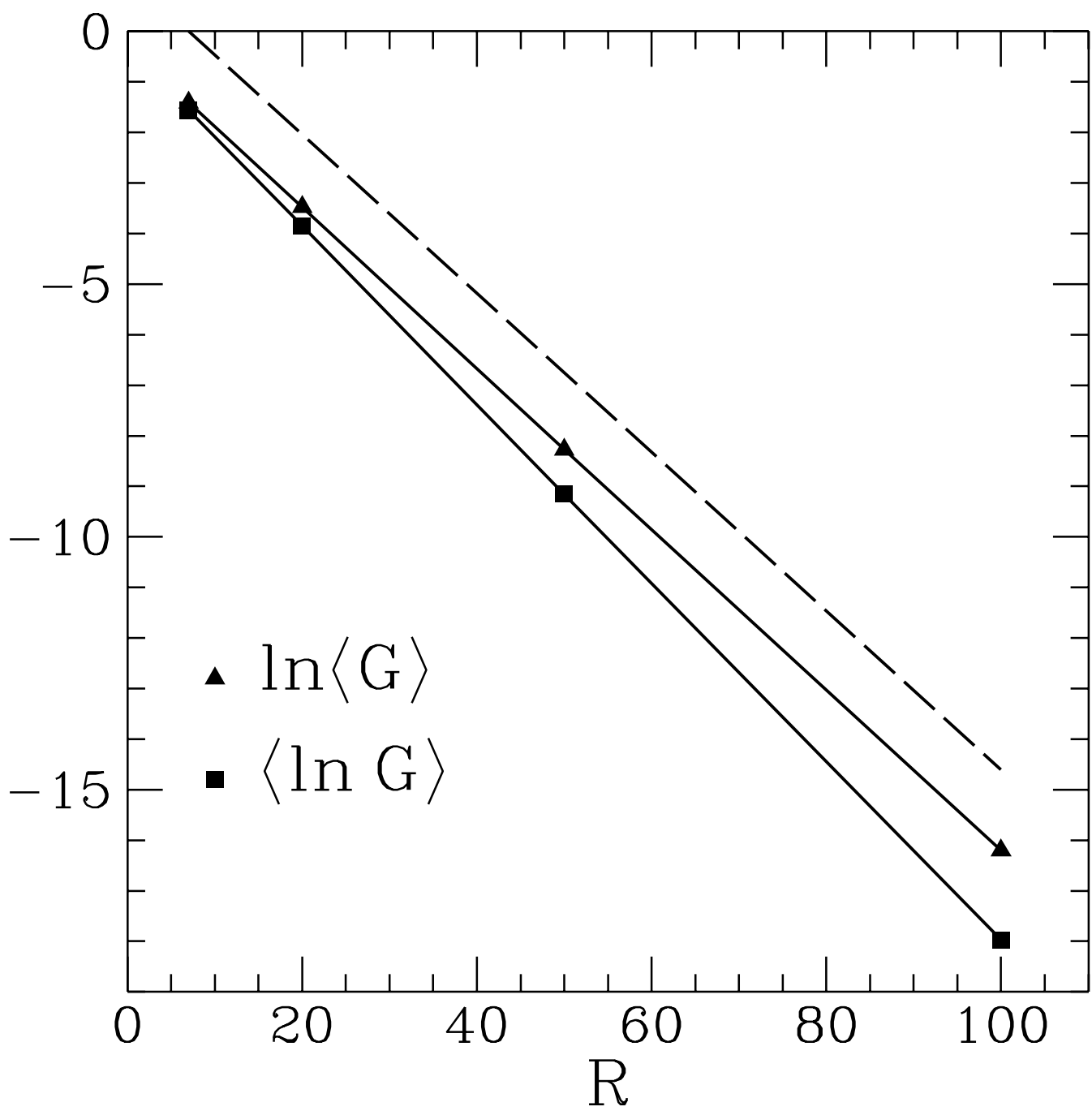
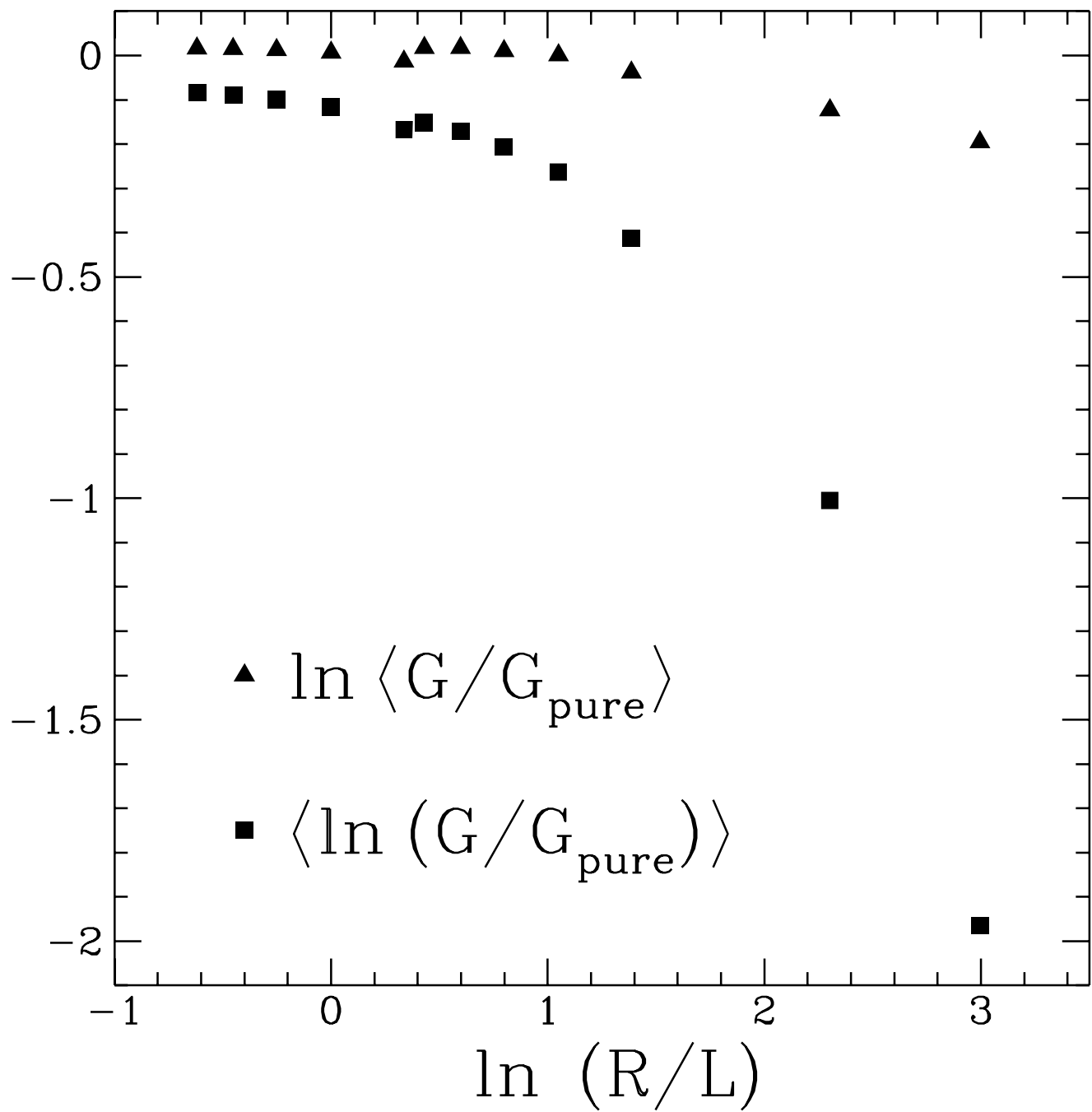
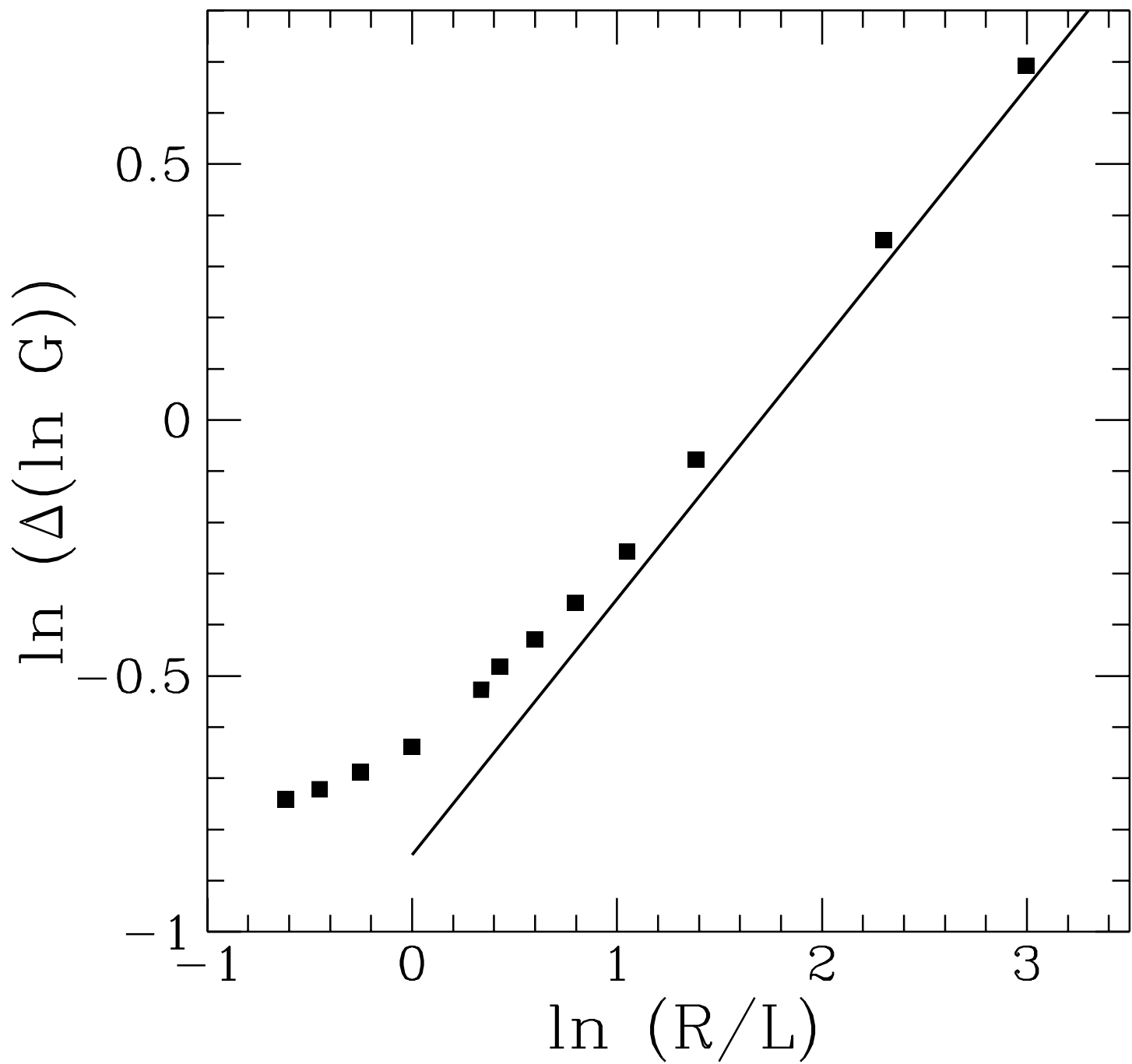
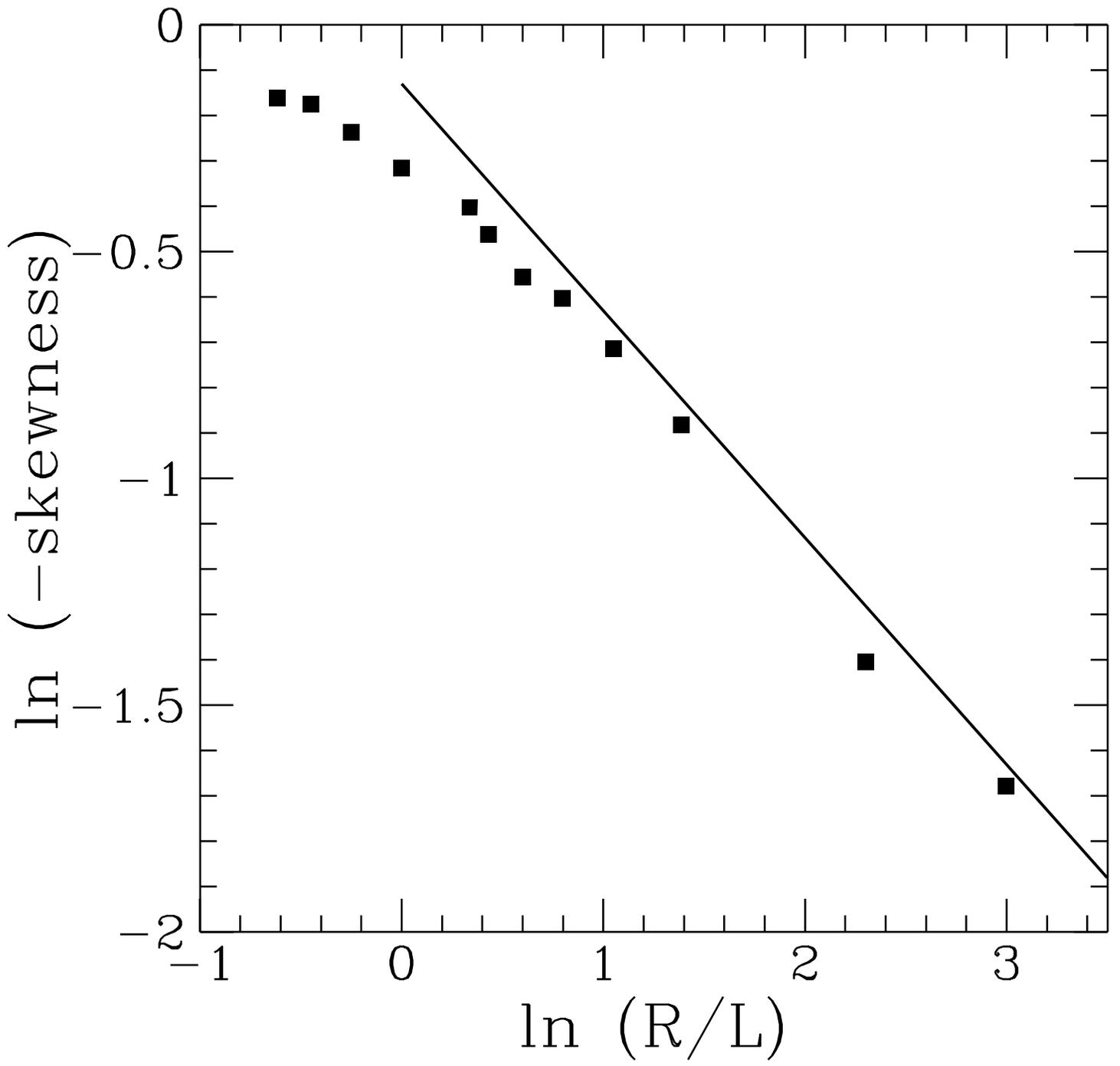
\end{document}